\def\spose#1{\hbox to 0pt{#1\hss}}

\def\multleft#1{\hbox to size{\vbox {\halign {\lft{##}\cr #1}}\hfill}\par}
\def\multright#1{\hbox to size{\vbox {\halign {\rt{##}\cr #1}}\hfill}\par}

\def\today{\ifcase\month\or January\or February\or March\or April\or May\or
      June\or July\or August\or September\or October\or November\or December\fi
      \space\number\day, \number\year}

\def\H2{\hbox{H$_{2}$}}
\def\Lya{Ly$\alpha$~}

\documentclass[usegraphicx]{cvbmn2e}

\usepackage{times}
\usepackage{amssymb}
\usepackage{graphics}
\usepackage{epsfig}
\usepackage[dvips]{color} 
\usepackage{multicol}
\usepackage[authoryear]{natbib2}
\usepackage{aas_macros}
\include{defn}

\begin{document}
\hsize=6truein

\title[The luminosity function of \Lya emitters at $2.3 < z<4.6$ from
integral-field spectroscopy]{The luminosity function of \Lya emitters
at $2.3 < z <4.6$ from integral-field spectroscopy\thanks{Based on
observations performed at the European Southern Observatory, Chile
[programme ID: 71.B-3015(A)].}}
\author[Caroline van Breukelen, Matt J.Jarvis \& Bram P. Venemans]
{Caroline van Breukelen$^{1,2}$\thanks{Email: cvb@astro.ox.ac.uk},
Matt J.~Jarvis$^1$\thanks{Email: mjj@astro.ox.ac.uk} \& Bram P. Venemans$^2$ \\
\footnotesize
$^1$Astrophysics Department, Keble Road, Oxford, OX1 3RH, United Kingdom. \\
$^2$Sterrewacht Leiden, PO Box 9513, 2300 RA, Leiden, The Netherlands\\}
\maketitle

\begin{abstract}
We have used VIsible MultiObject Spectrograph Integral-Field Unit
(VIMOS-IFU) observations centred on a radio galaxy at $z=2.9$ to
search for \Lya emitters within a comoving volume of $\sim 10^4$
Mpc$^3$. We find 14 \Lya emitters with flux $> 1.4
\times 10^{-20}$ W~m$^{-2}$, yielding a comoving space density of 
$0.0018^{+0.0006}_{-0.0005}$ Mpc$^{-3}$. We fit a Schechter luminosity
function which agrees well with previous studies both at similar
redshift ($z \sim 3.4$) and higher redshift ($z \sim 5.7$). We
therefore find no evidence for evolution in the properties of
\Lya emitters between $3 < z < 6$, although our sample is small.
By summing the star-formation rates of the individual \Lya emitters we
find a total cosmic star-formation rate density of $\rho_{SFR} =
6.7\pm 0.5\times10^{-3}$ M$_{\odot}$yr$^{-1}$Mpc$^{-3}$. Integrating
over the luminosity function for the combined \Lya surveys at $z \sim
3.4$ and accounting for the difference in obscuration between the \Lya
line and the UV-continuum yields an estimate of $\rho_{SFR} \sim
2.2\times10^{-2} $ M$_{\odot}$yr$^{-1}$Mpc$^{-3}$, in line with
previous multi-colour and narrow-band surveys of high-redshift
star-forming galaxies.

The detection of high-redshift emission-line galaxies in our
volumetric search shows that the unique capabilities of wide-field
integral-field spectroscopy are well suited in searching for
high-redshift galaxies in a relatively unbiased manner.
\end{abstract}

\section{INTRODUCTION}
One of the most important questions in cosmology is how the Universe
evolved into what we see today. It is believed that structure forms
from the initial density perturbations in the dark matter distribution
which collapse under gravitational instability
\citep{Eggen62, Sandage70, Peebles71, Press74, White78}. The
gravitational attraction of the dark matter overdensity causes gas to
accumulate and cool, thus enabling star formation
\citep{White78}. Hierarchical models of galaxy formation
postulate that these sub-galactic clumps subsequently merge to form
larger galaxies. Studying star-forming galaxies as a function of both
luminosity (or star-formation rate) and redshift is therefore
essential to obtain insight into the formation history of galaxies.

As galaxies form and evolve they also influence their environment. 
The most obvious way in which this happens is that they are able to
ionise the gas surrounding them. Observations of distant quasi stellar
objects (QSOs) exhibit absorption shortward of the \Lya line
\citep{Gunn65}, which indicates that the reionisation of the Universe
was not yet complete at $z > 6$ \citep[e.g.][]{Becker01,
Djorgovski01}. The vital question remains: which objects are
responsible for this process? Again analysing the properties of
primeval galaxies can provide an answer as the radiation emitted by
the young OB-stars could be energetic enough to ionise the
intergalactic medium \citep[see e.g.][]{Madau99}.

To detect primeval galaxies and determine the cosmic star-formation
rate density, two different types of object are often targeted. One
comprises the Lyman Break Galaxies, which are star-forming galaxies
characterised by a strong break in their spectrum shortward of the
Lyman limit at 912\AA\ and the hydrogen \Lya line at 1216\AA. The
other type is the focus of this paper: the \Lya emitters,
identified by their prominent emission in the \Lya line. Lyman Break Galaxies
have been discovered in huge numbers over the past decade using
multi-colour surveys \citep[e.g.][]{Steidel99}. Using this technique
several surveys, probing different ranges in redshift, indicate a
roughly constant star-formation rate density between $1 < z < 5$
\citep{Connolly97, Madau98, Steidel99, Iwata03}. However, at $z > 6$
the star-formation rate density seems to decline \citep{Stanway03,
Bunker04} possibly indicating that the source of reionisation may not
be the star-forming galaxies we see at $z < 6$.  Searches for \Lya
emitters are complementary in this goal as these surveys can probe
galaxies with relatively very weak continuum emission - undetected by
the Lyman Break technique.

Several techniques have been used to observe high-redshift \Lya
emitters: (i) slitless spectroscopy \citep[e.g.][]{Koo80}, (ii)
slit-spectroscopy \citep[e.g.][]{Lowenthal90, Thompson92}, and (iii)
narrow-band imaging \citep[e.g.][]{Hu97, Cowie98}. Each method has its
pros and cons. Slitless spectroscopy allows a
large volume to be probed, but needs very long integration times to
achieve an acceptable flux limit. Slit-spectroscopy can achieve a good
flux limit with shorter integration times but unfortunately the
surveyed volume is often very small and multi-colour imaging
observations are usually needed to preselect candidates. Narrow-band
imaging compares broad-band images to images taken with a filter of
very narrow bandwidth (e.g. $\Delta \lambda \sim 60$\AA). Objects that
are significantly brighter in the narrow-band image are likely to have
a strong emission line at the observed wavelength of the narrow-band
filter. This way large areas can be searched for \Lya emitters. This
type of survey has been the most successful in detecting \Lya emitters
since the late 1990s when the sensitivity of the surveys increased
considerably due to the onset of 8-metre and 10-metre class telescopes
\citep[e.g.][]{Hu97, Cowie98}.

With this leap in collecting area, detection of \Lya emitters through
narrow-band imaging has frequently been used to determine the cosmic
star-formation rate density from $z \approx 3 - 7$
\citep[e.g.][]{Kudritzki00, Fujita03, Kodaira03}. Nevertheless
narrow-band imaging suffers from one major drawback: due to the small
FWHM of the narrow-band filter, the surveys sample a very narrow
redshift range, and thus a shallow volume. Hence it is impossible to use
a single narrow-band survey to determine the Ly$\alpha$ density and
luminosity function throughout a large redshift range. 

In this paper we aim to improve upon these limitations by searching
for Ly$\alpha$ emitters with integral-field spectroscopy. An 
Integral-Field Unit (IFU) samples an area on the sky where every pixel contains
a spectrum. This then allows a volume of the Universe to be probed
with just a single observation. Thus depending on the wavelength range
probed we are able to study the comoving space density and
properties of Ly$\alpha$ emitters over a large redshift range.

We have used the VIsible MultiObject Spectrograph Integral-Field Unit
\citep[VIMOS-IFU, ][]{LeFevre03} on the Very Large Telescope to search
for \Lya emitters within a volume of $\sim 10^4$ Mpc$^3$. In section 2
we describe our observations and the data reduction and section 3
details the selection procedure used to find \Lya emitters in our
data. The results are presented in section 4; section 5 discusses our
findings and in section 6 we provide a summary of our
conclusions. Throughout this paper we assume $\Omega_{\rm matter}=0.3$,
$\Omega_{\Lambda}=0.7$, and $H_0=71$ km s$^{-1}$Mpc$^{-1}$, unless
stated otherwise.

\section{OBSERVATIONS AND DATA REDUCTION}
\begin{figure}
\begin{center}
\includegraphics[width=10cm]{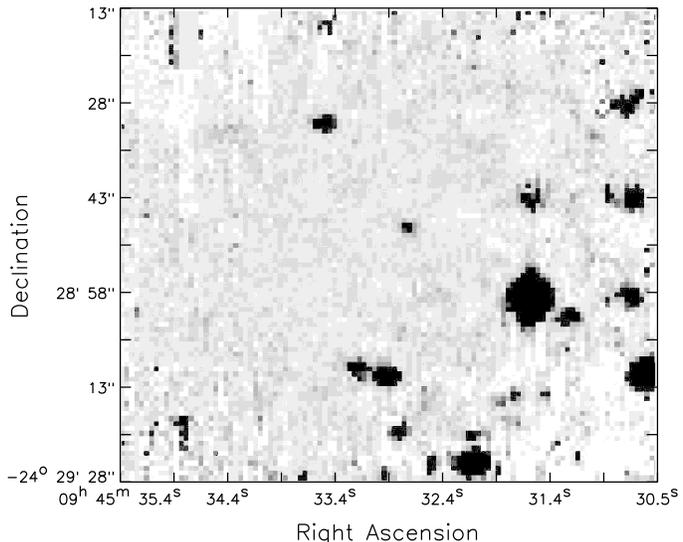}
\caption{\small The two-dimensional image of our IFU field of view,
integrated from 4050 \AA\ to 6800 \AA.}
\label{2Dimage}
\end{center}
\end{figure}

All of our observations were made on the nights of 29 April 2003 to 2
May 2003 on the 8.2-metre Melipal telescope at the ESO Paranal
Observatory in Chile. We used the VIMOS-IFU which consists of 6400
fibres coupled to microlenses. Each fibre produces a spectrum of a
small region on the sky; these spectra can be integrated to construct
a `broad-band' image, as shown in Fig.~\ref{2Dimage}. Our observations
were made in low-resolution mode which corresponds to a resolution of
$R \approx 250$. The spectral wavelength range was 3500\AA\
$\rightarrow$ 7000\AA\ with a dispersion of 5.35 \AA\ per pixel. The
spatial sampling of the fibres in our set-up was 0.67 arcsec per
fibre, resulting in a square continuous field of view of $54 \times
54$~arcsec$^{2}$ with the dead space between the fibres comprising
less than 10\% of the fibre to fibre distance. The field is split up
into four quadrants, each connected with an EEV CCD detector
consisting of 2048 x 4096 pixels.

The observations were centred on MRC 0943-242, a powerful radio galaxy
at $z=2.92$ ($\alpha=9^h45^m32.7^s$, $\delta=-24^\circ28'49.7''$,
J2000). The total exposure time was 9 hours, split into 18 exposures of
30 minutes. In order to minimise the effects of bad pixels and
contamination by cosmic rays during data reduction, the consecutive
exposures were offset by 10 arcsec in both right ascension and
declination relative to the centre. During all four nights the seeing
varied between 0.6 and 0.8 arcsec.

The data were reduced using the VIMOS Interactive Pipeline Graphical
Interface \citep[VIPGI, ][]{Scodeggio04}. First the standard routines
for bias and flat-field correction are applied. Subsequently the
individual spectra on the science frames are separated from each
other. Each frame consists of 400 spectra, corresponding to one
quadrant of the IFU. The layout of the IFU fibres in the instrument
focal plane is used to calculate the position of the individual
spectra on the CCD frame. When the spectra are located, the wavelength
calibration is derived specifically for each fibre by identifying arc
lamp lines taken with the same set-up. The variations in the
transmission of the individual fibres are corrected for and a sky
estimate is calculated by averaging the individual spectra. The sky
spectrum is then subtracted from the individual spectra and cosmic
rays are removed using a 3$\sigma$ clipping algorithm on the data. The
spectra are flux calibrated using standard-star frames. To this end
the instrument sensitivity function is derived by summing the spectra
which contain flux from the standard star. This results in an accurate
relative flux calibration.  To obtain the absolute flux calibration,
we use a long-slit spectrum of the central radio galaxy taken with the
LRIS on the Keck Telescope. We measure the flux of the Ly$\alpha$ line
by integrating the flux in the LRIS-spectrum and construct a spectrum
of the same object with the VIMOS-IFU data by summing the fibres along
the long-slit orientation. This allows us to bootstrap the
spectrophotometry. Comparisons of spectra and V-band photometry of
several objects in the field show that the final flux calibration is
good to $\sim 20$~per cent. The effect of the dead space between the
fibres on the error in the flux measurement is negligible as it is
inherently small and has been even further reduced by the dithering of
our single exposures.

The science frames are median combined to remove any remaining cosmic
rays and a fringe correction is applied by median combining
the fringe frames. The final data cube consists of 12100 combined
spectra, as a result from our enlarged field of view due to the
dithering of the individual exposures. These 12100 spectra correspond
to a two-dimensional image of $110 \times 110$ pixels with a field of
view of $1.2' \times 1.2'$ (see Fig.~\ref{2Dimage}).

\section{DATA ANALYSIS}
\subsection{The sensitivity of the data}\label{sec:sensitivity}
To determine the sensitivity of the data a wavelength range is
selected where the flux calibration is good and the spectra are clear
of skylines. At both ends of the spectra the flux calibration is poor,
so we discard the regions at $3500$\AA\ $ < \lambda < 4200$\AA\ and
$6800$\AA\ $ < \lambda < 7000$\AA. Between $\lambda=5500$\AA\ and
$\lambda=6550$\AA\ three skylines dominate the spectra; this region is
also omitted from the determination of the sensitivity function. In
each spectrum we fit a polynomial to the usable wavelength range to
facilitate calculation of the Root Mean Square (RMS) of the noise
($\sigma$). The fit is subtracted from the data and $\sigma$ is
calculated. As the presence of actual features in the spectrum would
alter the value of the noise, data points deviating more then
$3\sigma$ from the fit are discarded. Subsequently the RMS of the
noise is recalculated. This procedure results in a measurement of the
overall RMS of the noise in each spectrum.
 
\begin{figure}
\begin{center}
\includegraphics[width=9cm]{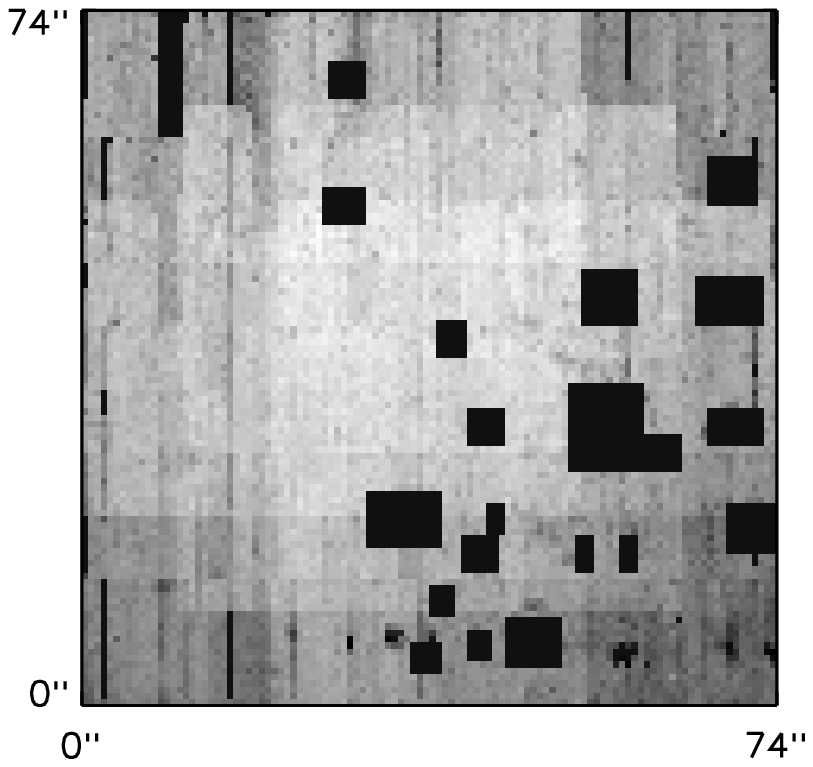}
\vspace{0.5cm}
\includegraphics[width=9cm]{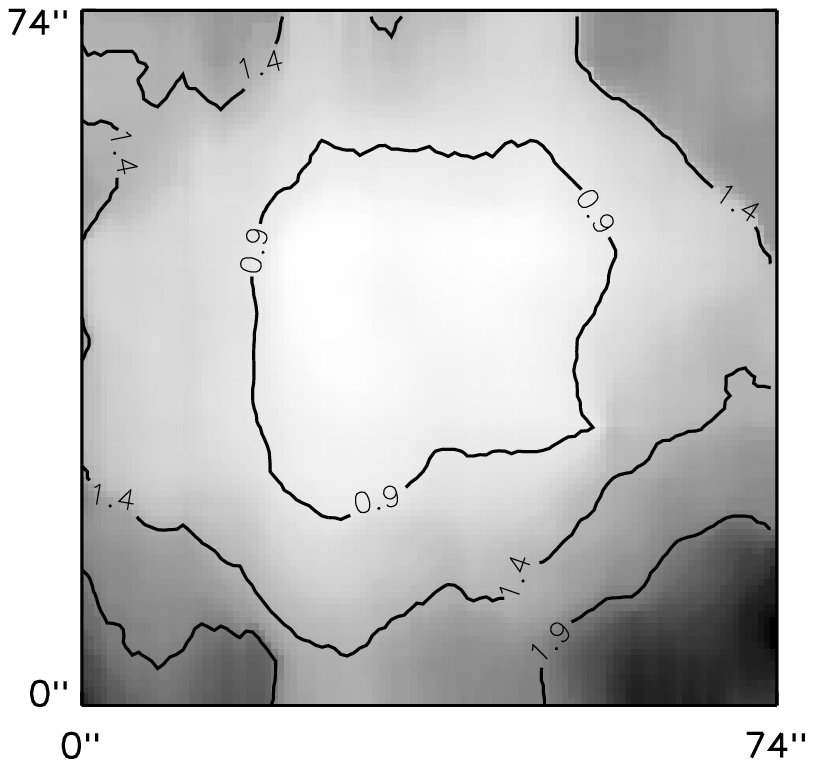}
\caption{\small The sensitivity function over our field of view. The
lighter regions correspond to better sensitivity. Top: pixel to pixel
sensitivity function; regions with bright objects are set to
zero. Bottom: smoothed sensitivity function to show the effects of the
dithered single exposures. The $5\sigma$ sensitivity varies
between $6.8 \times 10^{-21}$~W~m$^{-2}$pix$^{-1}$ (white) to $3.1
\times 10^{-20}$~W~m$^{-2}$pix$^{-1}$ (black). The contours shown are
at 0.9, 1.4, and $1.9\times^{-20}$~W~m$^{-2}$pix$^{-1}$.}
\label{sensitivity}
\end{center}
\end{figure}

Visual inspection of random spectra with the continuum fit and noise
estimate confirm that the measurements are accurate.  The method does
not work well in fibres where there are bright objects, e.g. stars or
nearby galaxies, due to the large number of emission and/or absorption
lines in the spectrum. However, to search for distant Ly$\alpha$
emitters we want to avoid those regions of the sky with bright nearby
objects. Therefore the sensitivity determination described above is
well suited to our needs as it yields an accurate value for the noise
in spectra that do not sample luminous sources. To be able
to exclude regions with bright objects the noise value in the
corresponding fibres is set to zero. Inspection of the spectra of
these objects reveals them to be nearby galaxies, stars, a QSO at
$z=1.8$ \citep[see][]{Jarvis05} and a galaxy at $z=0.62$, identified
through its [OII] emission.  In Fig.~\ref{sensitivity}
the sensitivity function over the two-dimensional image is shown. Note
that, especially in the smoothed image, the effects of dithering the
exposures are evident. The noise varies from $\sigma = 1.4\times
10^{-21}$~W~m$^{-2}$pix$^{-1}$ to $6.2\times
10^{-21}$~W~m$^{-2}$pix$^{-1}$ over the data cube.

\begin{table*}
\caption{\small Observed \Lya emitters with their observational
properties. The typical errors on the quoted positions are 0.3" and
reflect the size of the IFU-fibres. The wavelength, flux, and their
errors are calculated by a repeated Gaussian fit to the emission lines
as described in section~\ref{sec:emitters}. A star symbol ($*$)
following the candidate number signifies an extended source (i.e. the
emission line is spread over more than one IFU fibre). All
luminosities $L_{\rm line}$ quoted assume that the line lies at the
redshift $z$, given in column~6.  }
\hspace{-1.5cm}
\begin{tabular}{r c c c c c c c}
\hline \hline
Nr. & RA (J2000) & Dec & $\lambda$ & F$_{line}$ & z & L$_{line}$ \\
& $\rm ^h$~~~$\rm ^m$~~~$\rm ^s$ & $^{\circ}$~~~~$'$~~~~$''$& \AA & $10^{-20}$ W m$^{-2}$&  & $10^{35}$ W \\
\hline
532*  & 9:45:30.64 & -24:28:26.9 & 5752$\pm$1 & 3.0$\pm$0.7 & 3.7315$\pm$0.0008 & 4.1$\pm$0.9 \\  
1315* & 9:45:30.95 & -24:28:18.2 & 5140$\pm$1 & 6.7$\pm$1.8 & 3.2281$\pm$0.0008 & 6.4$\pm$1.7 \\ 
1702* & 9:45:31.13 & -24:28:53.7 & 5699$\pm$1 & 2.7$\pm$0.8 & 3.6879$\pm$0.0008 & 3.6$\pm$1.1 \\  
2919  & 9:45:31.62 & -24:28:49.0 & 5437$\pm$1 & 1.4$\pm$0.4 & 3.4724$\pm$0.0008 & 1.6$\pm$0.5 \\ 
3210* & 9:45:31.76 & -24:29:15.1 & 4676$\pm$1 & 5.5$\pm$1.0 & 2.8464$\pm$0.0008 & 3.9$\pm$0.7 \\  
3510  & 9:45:31.85 & -24:28:21.5 & 6601$\pm$1 & 2.1$\pm$0.5 & 4.4299$\pm$0.0008 & 4.3$\pm$1.0 \\ 
5644* & 9:45:32.74 & -24:29:05.8 & 4766$\pm$1 & 7.3$\pm$1.4 & 2.9205$\pm$0.0008 & 5.5$\pm$1.1 \\
6164  & 9:45:32.96 & -24:29:25.9 & 4341$\pm$1 & 3.9$\pm$0.8 & 2.5709$\pm$0.0008 & 2.1$\pm$0.4 \\
6537  & 9:45:33.10 & -24:28:57.1 & 6619$\pm$1 & 1.5$\pm$0.3 & 4.4447$\pm$0.0008 & 3.1$\pm$0.6 \\ 
8060  & 9:45:33.72 & -24:29:08.4 & 4861$\pm$2 & 1.7$\pm$0.6 & 2.9986$\pm$0.0016 & 1.4$\pm$0.5 \\
8599  & 9:45:33.95 & -24:29:15.8 & 4280$\pm$1 & 1.8$\pm$0.5 & 2.5207$\pm$0.0008 & 0.9$\pm$0.3 \\
8827  & 9:45:34.03 & -24:29:10.5 & 4234$\pm$2 & 2.1$\pm$0.7 & 2.4829$\pm$0.0016 & 1.1$\pm$0.4 \\
8893  & 9:45:34.03 & -24:28:26.2 & 4611$\pm$1 & 2.3$\pm$0.5 & 2.7930$\pm$0.0008 & 1.5$\pm$0.3 \\
10913 & 9:45:34.88 & -24:29:13.1 & 4232$\pm$1 & 2.3$\pm$0.6 & 2.4812$\pm$0.0008 & 1.2$\pm$0.3 \\ 
\hline \hline
\end{tabular}

\label{lines}
\end{table*}

\subsection{Selection of line emitters}\label{sec:selection}
With a dispersion of ~5.35\AA\ per pixel and a resolution of
$R \approx 250$, the FWHM of the emission lines will
generally be only about 3 pixels as we expect the typical \Lya emitter
to have an unresolved \Lya line at this resolution. As the lines may
well appear narrower due to the noise, we select peaks with a maximum
value over $4\sigma$ above the previously determined fit to the
continuum level and a 2-pixel integrated flux greater than $5\sigma$.
Taking a lower limit than 4$\sigma$ for the peak value of the line
leads to too much confusion with peaks in the noise; furthermore the
detection completeness of the low-flux lines would be very low. The
integrated 5$\sigma$ flux limit will result in the detection of
emission lines with a lower limit to the flux varying over the whole
field between 1.0$\times 10^{-20}$~W~m$^{-2}$ and 4.4$\times
10^{-20}$~W~m$^{-2}$ for an unresolved emission line, as derived from
the sensitivity function. We only investigate the wavelength ranges
$4050$\AA\ $<
\lambda < 5520$\AA, $5625$\AA\ $<
\lambda < 5850$\AA, $5950$\AA\ $< \lambda < 6130$\AA, and $6550$\AA\
$< \lambda < 6800$\AA\ to avoid skylines and distortion by bad flux
calibration. After selecting all peaks in the spectra that satisfy the
imposed criteria, the peaks that are clearly caused by bad pixels,
poor flux calibration or other irregularities in the data are
discarded. We verify the selection procedure by performing a number of
tests. First, the same selection process is run again with a smoothed
sensitivity function (see bottom Fig.~\ref{sensitivity}) instead of
the pixel-to-pixel values (top Fig.~\ref{sensitivity}). In this way we
check if any candidates are selected or rejected due to errors in the
noise determination in individual fibres; we also run this test
without fitting the continuum, to make sure there are no
line candidates selected or discarded through an erroneous fit. The
spectra of the remaining line candidates are then inspected by eye. To
this point the RMS of the least noisy parts of the spectrum have been
used to identify candidate line emitters. However, some parts of the
spectrum are noisier, therefore the local RMS of the noise around the
line candidate is determined to check if the peak value is still above
4$\sigma$. We then calculate the integrated flux over the two highest
pixels to ensure that it is more than 5 times the RMS of the
integrated local noise.\\
\indent Bad pixels and cosmic rays affect the final image even in the data
cube consisting of the combined frames. Moreover first order
superpositions are not well removed during the data reduction nor are
the reduced frames always clear of sky line residues. Therefore we
construct two more data cubes, comprising 8 and 10 original exposures
respectively, and calculate at what level the line should be detected
if it is real. If the line is clearly present in one cube, but not in
the other, it is likely to be an artefact seen in only one or a few
original frames. Subsequently for each line candidate all the single
exposures are visually inspected for artefacts. The final check is to
see if the line could be the result of crosstalk between fibres, which
is caused by overflowing of the light from a fibre into a neighbouring
fibre on the CCD. These neighbouring fibres do not necessarily
correspond to adjoining positions on the sky as there is a complicated
mapping of fibres from the IFU-head onto the CCD. If a line candidate
passes all these tests, it is declared to be a real emission
line. After all of these consistency checks we find 15 emission-line
galaxies.

\section{RESULTS}
\subsection{Properties of the line emitters}\label{sec:emitters}
Our IFU data facilitates examination of both the spectral and spatial
properties of the 15 line emitters we have found. The spectra of the
14 candidate \Lya emitters are shown in Fig.~\ref{spectra}. The other
source is a narrow-line active galaxy which is discussed fully in
Jarvis et al. (2005). 

\begin{figure*}
\begin{center}
\includegraphics{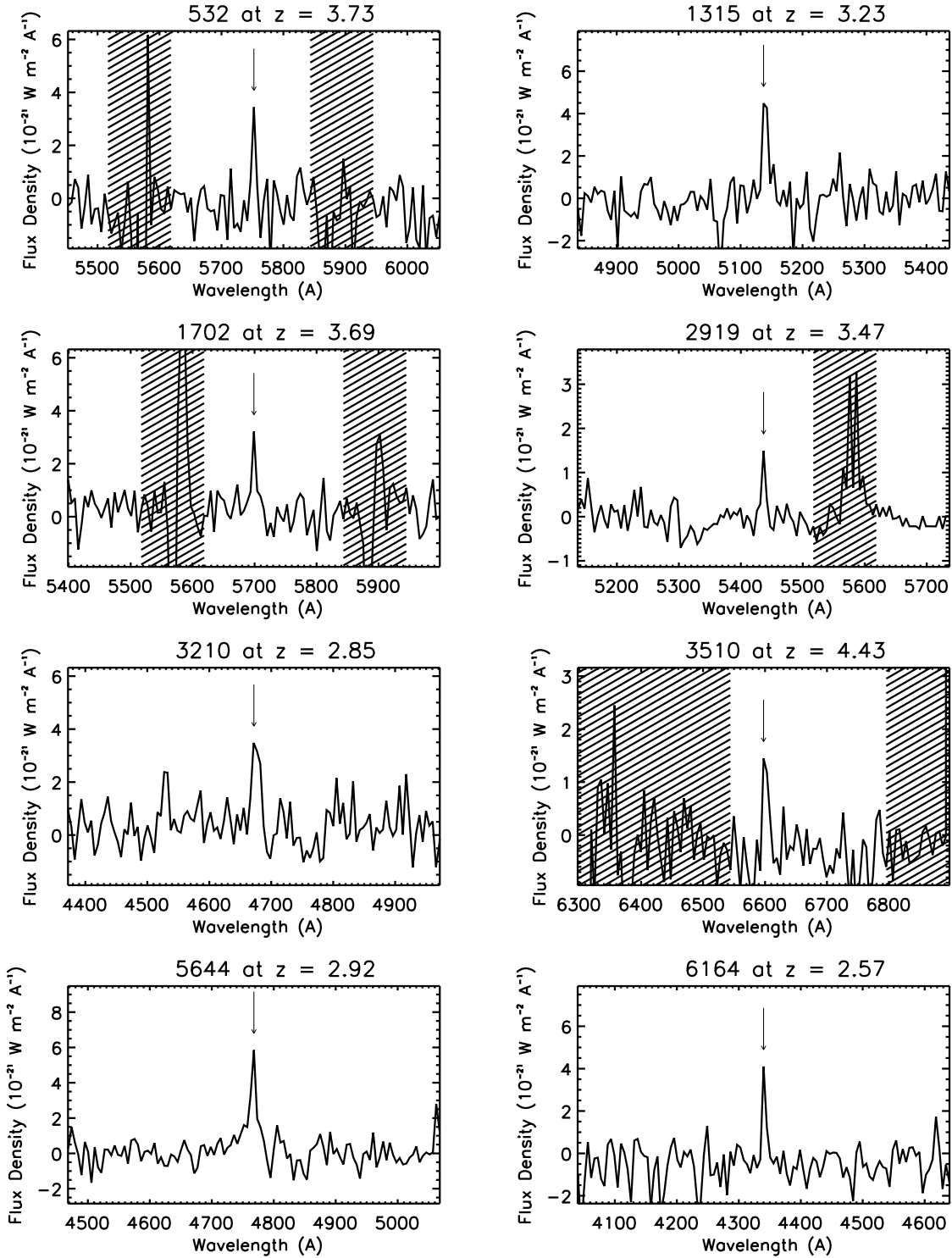}
\end{center}
\caption{\small Spectra of the identified line emitters. The positions
of the emission lines are denoted by an arrow.  The regions that were
excluded from our analysis are shaded; at these wavelengths skylines
or bad flux calibration dominate the spectra. The redshifts all assume
that the emission line is \Lya.}\label{spectra}
\end{figure*}
\addtocounter{figure}{-1}
\begin{figure*}
\begin{center}
\includegraphics{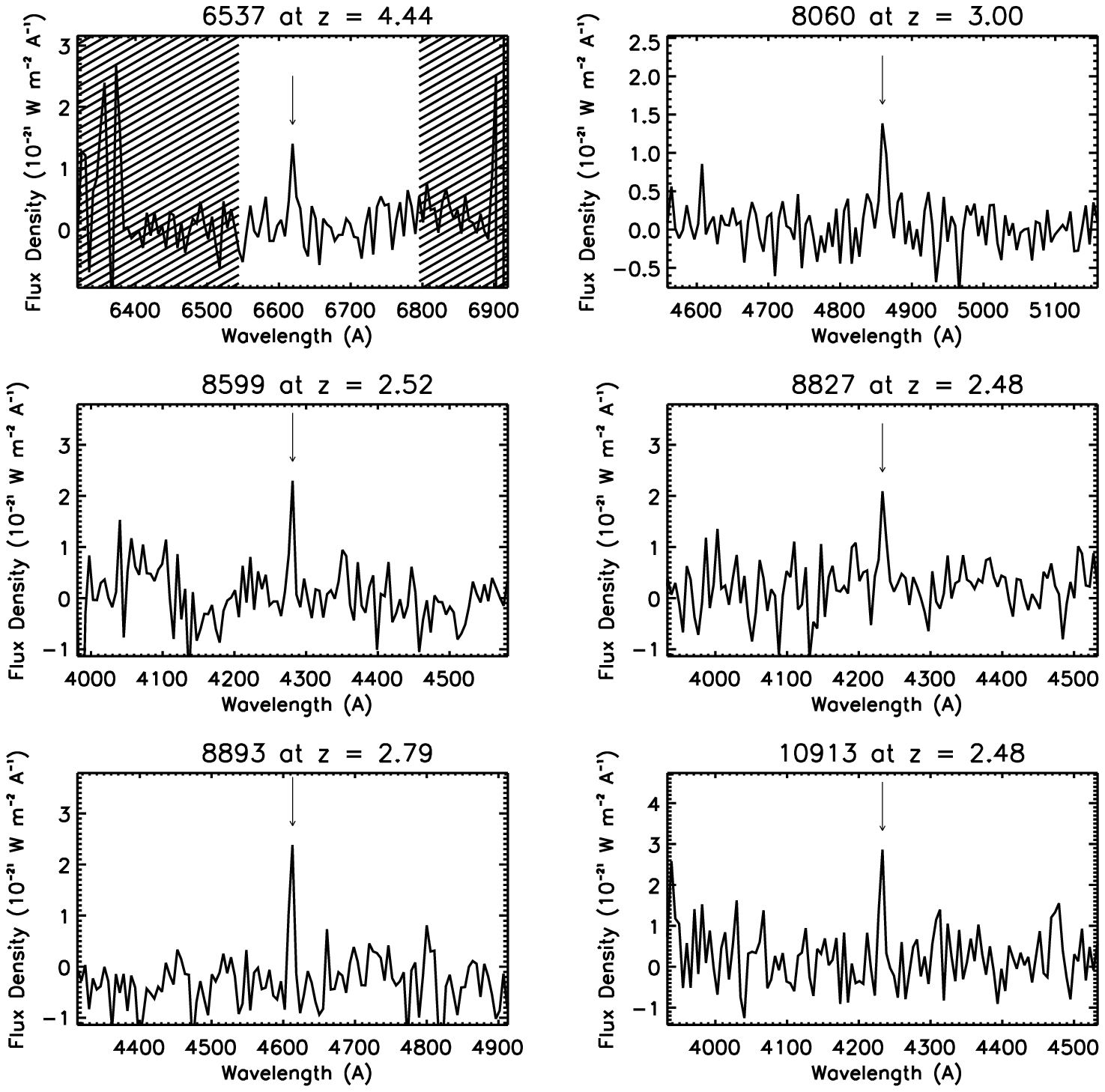}
\end{center}
\caption{cont.}
\end{figure*}

For each candidate \Lya emitter we measure the total flux of the
emission line by integrating over the spectra of all fibres sampling
the source. The flux of each emission line is measured by fitting a
Gaussian profile to the emission line. To determine the uncertainty in
the line profile, we recalculate our fit 1000 times, each time adding
to the spectrum random noise with an RMS corresponding to the local
noise in the spectrum at the position of the emission line. The flux
and the wavelength of the line are taken to be the mean of the 1000
calculations. The standard deviations correspond to the $1\sigma$
uncertainties on the flux and wavelength. The results of these fits
can be found in Table~\ref{lines}.

\begin{figure}
\begin{center}
\includegraphics[width=10cm]{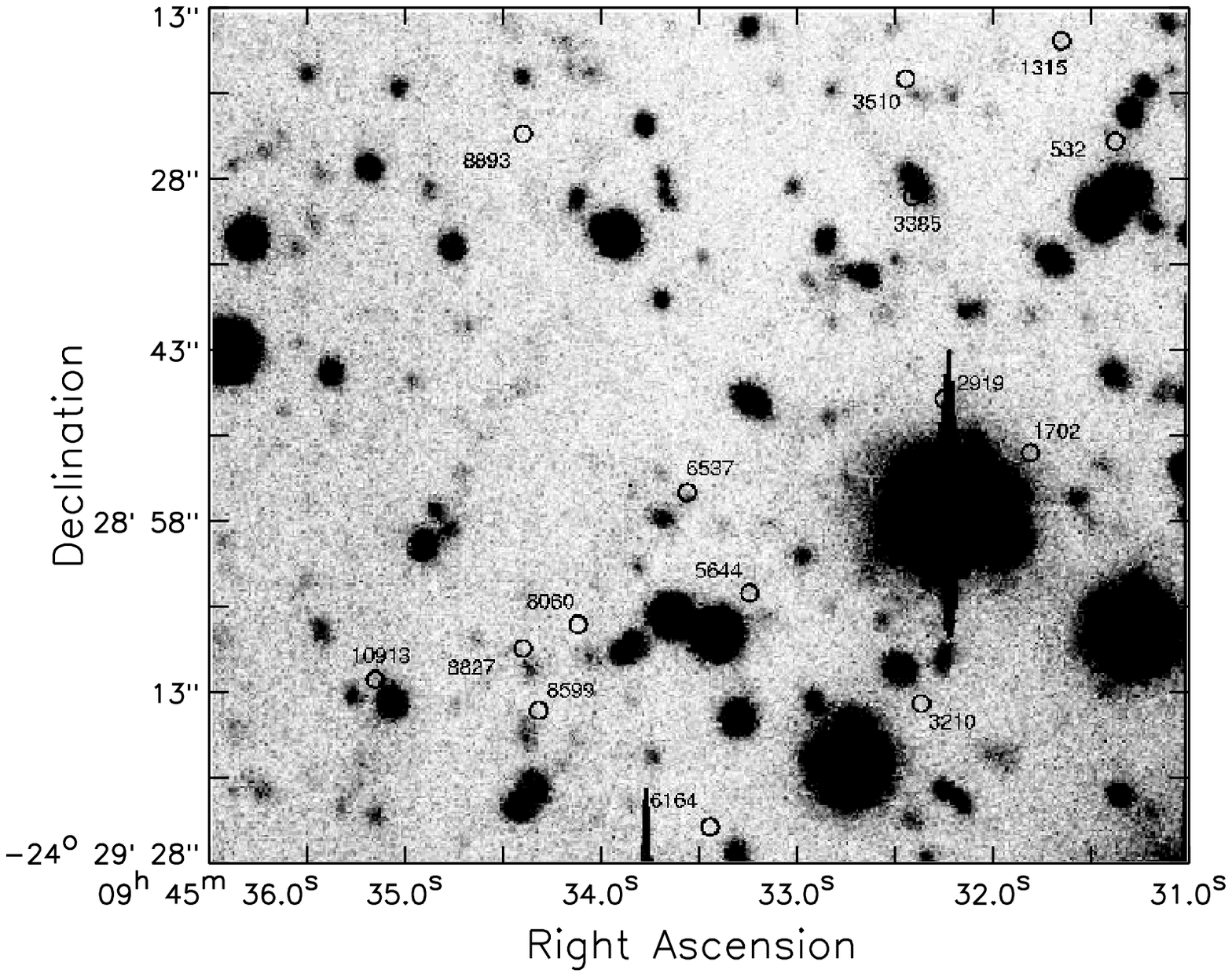}
\caption{\small Keck V-band image of our field of view. The positions
of the line emitters are marked by a circle the size of an IFU
fibre. The accuracy in RA is $\sim 0.3"$; in Dec the error
increases with distance from the centre from $0.3"$ to up to $\sim 1"$. This
compromises the certainty of the identification of object 3510, 8599,
and 8827.}
\label{Vimage}
\end{center}
\end{figure}

With broad-band images of this field\footnote{These are (1) a B-band
image with a 2$\sigma$ limiting magnitude of $m_{B,limit}=28.4$, taken
with the Fors2 instrument on the VLT, (2) a V-band image with
$m_{V,limit}=28.2$, taken with the LRIS camera on the Keck Telescope
and (3) an I-band image with $m_{I,limit}=25.1$, also taken with LRIS
on the Keck Telescope.}, we derive the continuum level of the
candidate line emitters by measuring (or placing a lower limit on)
their broad-band magnitude. The sources are taken to be detected if
the flux within the IFU-fibre aperture is larger than $2\sigma$ above
the background level. The V-band image is shown in Fig.~\ref{Vimage};
the locations of the line-emitters are indicated by circles which
reflect the errors on the positions as quoted in Table~\ref{lines}. To
verify the correctness of the identification of line emitters with
objects in the broad-band images we compare the coordinates of six
bright objects that are easily visible in the IFU data cube as well as
the broad-band image. The mean difference between the two positions
for each object is $\sim 0.3"$ in RA. In declination however the error
varies from $\sim 0.3"$ in the centre of the field to $\sim 1"$ at the
edges, and we return to this point in section~\ref{sec:id}.

\subsection{Identification of the emission lines}\label{sec:id}
It is possible not all our objects are actually Ly$\alpha$ emitters at
high redshift, such that some may be low-redshift interlopers. Within
our wavelength range any emission line is most likely to be either
\Lya$\lambda1216$, [OII] $\lambda3727$,
[OIII] $\lambda5007$, or H$\alpha~\lambda6563$.  For each spectrum we
inspect the plausibility of these four cases. The redshift of the
object is calculated for each possible line and subsequently the
redshifted wavelength of other emission lines are inspected by
checking whether the signal at these positions is greater than
$2\sigma$ above the local noise, which would indicate a possible
detection. None of the spectra were found to exhibit any other lines
under this criterion, which rules out the [OIII] $\lambda5007$ and
H$\alpha~\lambda6563$ possibilities.  However confusion with the
[OII] $\lambda 3727$ line may still occur, since we would not expect to
see other bright emission lines in our spectra if this is the observed
line. The most distinguishing characteristic of the [OII] $\lambda
3727$ line is the fact that it is actually a doublet with lines at
$\lambda=3726$\AA\ and $\lambda=3728$\AA; unfortunately due to our low
resolution we cannot resolve this feature. We investigate how likely
our lines are to be [OII] by calculating how many [OII] emitters we
would expect to find in our volume according to Hogg et al. (1998),
who carried out a survey at $z \sim 1$. For our observed wavelength
range the [OII] emitters within our field of view would be at $0.087 <
z < 0.825$, corresponding to a volume of 214 Mpc$^3$ (here we use
$\Omega_{\rm M}=0.3$, $\Omega_{\Lambda}=0$, and $H_0=100$ km
s$^{-1}$Mpc$^{-1}$ to facilitate comparison with the [OII] survey of
Hogg et al., 1998). Using the [OII] luminosity function combined with
the sensitivity function derived from our data, we expect to detect 3
[OII] emitters in our volume. Indeed we have found one [OII] emitter
at $z=0.62$ (see section~\ref{sec:sensitivity}) among the bright
extended objects that we have excluded from our selection
procedure. This galaxy could easily be identified as such by the clear
presence of the 4000\AA\ break in its continuum.

As derived from stellar synthesis models a Ly$\alpha$ line,
originating from a young star-forming region, has a typical rest-frame
equivalent width (EW$_0$) of 100-200\AA,
\citep{Charlot93}. Surveys for Lyman Break
Galaxies however often observe far lower equivalent widths for the
\Lya line. At $z=3$ \citet{Shapley03} find that only half of the Lyman
Break Galaxies with observed \Lya emission show an EW$_0$ of $>
20$\AA\ and very few objects have EW$_0> 100$\AA. Also at $z=6$ Lyman
Break Galaxies usually show a detected EW$_0$ of $\sim 20-30$\AA\
\citep{Stanway04}. Nevertheless observations of \Lya emitters at high
redshift show that many of these objects show a \Lya EW$_0$ of $>
200$\AA\ \citep[e.g.][]{Malhotra02}. Studies of [OII] emitters on the
other hand show that their equivalent widths are considerably
lower. For example \citet{Jansen00} have carried out a survey of
nearby field galaxies, which resulted in equivalent width measurements
with a maximum of $\sim 80$ \AA\ for [OII]. Measuring the EW of our
emission lines thus provides a crucial test of whether the line is
likely to be Ly$\alpha$ or [OII]. Using the flux of the lines and the
(upper limit on) the continuum level, we calculate the (lower limit
on) equivalent widths of the lines; these are shown in
Table~\ref{EW}. For the objects detected in the band in which their
emission line falls, we can use the broad-band images to check whether
we can observe a spectral break due to hydrogen in the IGM. This is
impossible for the objects with emission lines in the B-band as we
have no bands at shorter wavelength. We have however one object
(no. 1315) with an emission line in V-band that is also detected in
that band; this object has $(B-V) > 1.3$. \citet{Dickinson98} imposes
a limit of $(B-V) > 1.5$ to select Lyman Break Galaxies, which means
that our B-band image is not deep enough to conclusively determine if
we see a strong spectral break, although the lower limit to the object's
colour is close to the selection criterion. The broad-band
identification for objects 3510, 8599, and 8827 are insecure due to
their proximity to faint objects in the broad-band images in regions of
the field where our astrometry becomes less secure ($\Delta \delta
\sim 1$~arcsec). Therefore we have calculated their EWs assuming both no
identification and an identification with the source found in close
proximity in the broad-band image.

As shown in Table~\ref{EW}, all but two of our unknown emission lines
would have rest-frame equivalent widths of several hundred \AA\ if
they were [OII]. The low-redshift survey for [OII]-emitters by
\citet{Jansen00} did not yield any [OII] emission lines with an EW$_0$
larger than 80\AA. In the survey at redshift $z \sim 1$ by
\citet{Hogg98} the highest detected EW$_{0}$ of [OII] is still below
110\AA. It is therefore highly improbable that any of our high
equivalent width emission lines are [OII]. Nevertheless there are two
emission lines, in objects 3510 and 6537, which could plausibly be
[OII], as the lower limit on their EWs would be $\sim119$\AA\ and
$\sim$ 84\AA\ respectively which is high but not impossibly
so. However it needs to be noted that this does not stem from a faint
line flux, but from the I-band magnitude limit which is much brighter
than those of the other broad-band images. This results in a lower EW
limit. Moreover the objects show no continuum or any other features in
their spectra, and it is therefore likely that these objects are
\Lya emitters too. Thus for the remainder of this paper we will assume
that all of our selected candidates are indeed \Lya emitters. The
inferred redshift and line luminosities for our objects are shown in
Table~\ref{lines}.

\begin{table}
\centering
\caption{The equivalent widths of the emission lines of our
candidate \Lya emitters. All values are lower limits unless preceded
by '=', which represents the actual EW instead of the lower limit.
EW$_{obs}$ is the observed equivalent width, EW$_{0,{\rm [OII]}}$ is
rest-frame EW assuming the observed line is [OII] and EW$_{0,{\rm
Ly}\alpha}$ is the rest-frame EW if the line is Ly$\alpha$. Quoted
errors are the result of the uncertainty in the measured flux,
magnitude, and redshift. Objects 3510, 8599, and 8827 have an insecure
identification in the broad-band images. For object 3510 this makes no
difference for the calculation of the EW as the nearby source is not
detected in I-band (in which wavelength range its emission line
falls). For objects 8599 and 8827 the EW is shown for both the case
where the nearby source is (1) unassociated and (2) associated with
the line emission.}
\begin{tabular}{r r r r}
\hline \hline
Nr. & EW$_{obs}$ & EW$_{0,OII}$ & EW$_{0,Ly\alpha}$ \\
& $10^2$ \AA & $10^2$ \AA & $10^2$ \AA \\ 
\hline
532      &  11$\pm$3     &   7$\pm$2   &  2.3$\pm$0.5  \\
1315     &= 13$\pm$5     & = 9$\pm$4   &= 3.0$\pm$1.2  \\
1702     &  10$\pm$3     &   6$\pm$2   &  2.0$\pm$0.6  \\
2919     &   7$\pm$2     &   5$\pm$1   &  1.7$\pm$0.5  \\
3210     &  12$\pm$2     &  10$\pm$2   &  3.2$\pm$0.6  \\
3510     & 2.1$\pm$0.5   & 1.2$\pm$0.3 & 0.39$\pm$0.09 \\
5644     &= 10$\pm$5     & = 8$\pm$4   &= 2.6$\pm$1.2  \\
6164     &  16$\pm$3     &  14$\pm$3   &  4.6$\pm$0.9  \\
6537     & 1.5$\pm$0.3   & 0.8$\pm$0.2 & 0.28$\pm$0.05 \\
8060     &   7$\pm$3     &   5$\pm$2   &  1.8$\pm$0.6  \\
8599$^1$ &   8$\pm$2     &   7$\pm$2   &  2.1$\pm$0.6  \\
8599$^2$ &= 2.9$\pm$0.6  &= 2.5$\pm$0.5&= 0.8$\pm$0.2  \\
8827$^1$ &   9$\pm$3     &   8$\pm$3   &  2.5$\pm$0.8  \\
8827$^2$ &= 1.9$\pm$0.3  &= 1.7$\pm$0.3&= 0.56$\pm$0.09 \\
8893     &  10$\pm$2     &   8$\pm$2   &  2.5$\pm$0.6  \\
10913    & 10$\pm$3      &   8$\pm$2   &  2.8$\pm$0.7  \\
\hline \hline
\end{tabular}

\label{EW}
\end{table}

\section{DISCUSSION AND CONCLUSIONS}

\subsection{The \Lya emitter luminosity function}

\begin{figure*}
\begin{center}
\includegraphics[width=16cm]{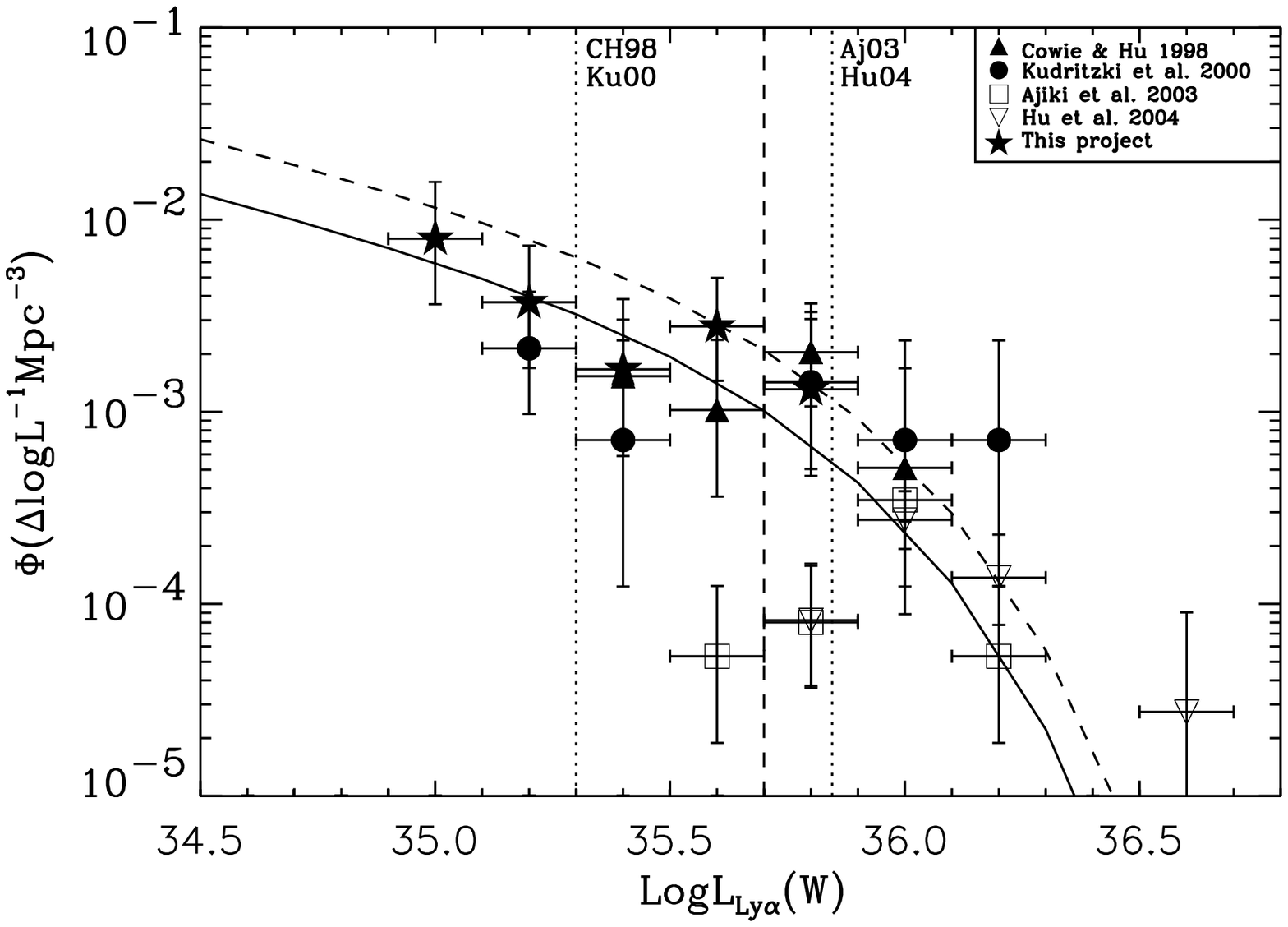}
\caption{\small The number density of Ly$\alpha$ emitters plotted
against the luminosity. The filled symbols mark surveys with an
average redshift similar to ours (\citet{Cowie98}: triangles,
\citet{Kudritzki00}: circles) and the open symbols stand for
surveys at redshift $z=5.7$ (\citet{Ajiki03}: squares, \citet{Hu04}:
upside-down triangles). Overplotted are two Schechter luminosity
functions: the solid line is the fit to all our data points and the
dashed line is the fit to our two highest luminosity data points and
those of the surveys at similar redshift with $L > 5 \times 10^{35}$
W (dashed horizontal line) to ensure completeness. The
dotted horizontal lines mark the detection limits of the surveys; for
Ajiki et al. (2003) this is the completeness limit.}
\label{discusplot}
\end{center}
\end{figure*}

Our field of view of 1.2' $\times$ 1.2' gives a solid angle
of 1.5 square arcmin. However, for our high-redshift
\Lya search we have omitted the regions which contain bright
objects. Accounting for this reduction in area due to the bright
sources, we are left with a solid angle of 1.36 square arcmin. If
we assume we are only looking at \Lya emitters, our observed
wavelength ranges (see section~\ref{sec:selection}) correspond to
redshift ranges of $2.33 < z < 3.54$, $3.63 < z < 3.81$, $3.89 < z <
4.04$, and $4.39 < 4.59$, which yield a total comoving volume of $7.8
\times 10^3$ Mpc$^3$.

The redshift range is large, thus the average flux limit over our
field of view of $1.8 \times 10^{-20}$ W m$^{-2}$ (see
Fig.~\ref{sensitivity}) corresponds to a luminosity limit which rises
significantly throughout the depth of our volume. If we calculate the
average luminosity limit over our field we find that it varies within
the extremes of our redshift range from $8 \times 10^{34}$~W at
$z=2.3$, to $4 \times 10^{35}$~W at $z=4.6$.

Due to the unique nature of our observations we are able to probe the
\Lya luminosity function across a wide range in luminosity, albeit
with a broad redshift bin. In Fig.~\ref{discusplot} we show the \Lya
luminosity function derived from our observations together with those
of \citet[][hereafter Ku00]{Kudritzki00} at $z\approx 3.1$, 
\citet[][hereafter CH98]{Cowie98} at $z = 3.4$, and \citet[][hereafter 
Aj03]{Ajiki03} and \citet[][hereafter Hu04]{Hu04} at $z\approx
5.7$. We have determined the volume of each of the luminosity bins by
calculating the volume within which a source of that luminosity can be
observed with our flux limit, which varies over the field of view
according to the sensitivity function derived in
section~\ref{sec:sensitivity}. Fig.~\ref{discusplot} shows that we
probe down to much lower luminosities than the narrow-band surveys at
$z > 3$; this stems from the fact that at the lower redshift limit of
our volume we have a significantly lower luminosity limit. The lack of
high-luminosity sources in our volume is caused by the small size of
the area surveyed. 

To determine if our luminosity function differs from
those found by the previous surveys at both similar redshift and higher
redshift, we fit a Schechter function to the data:

\begin{displaymath}
\Phi(L) dL = \Phi^*\Big(\frac{L}{L^*}\Big)^{\alpha}e^{-L/L^*}d\Big(\frac{L}{L^*}\Big),
\label{schechter}
\end{displaymath}
where $\Phi(L)$ is the comoving space density of sources per $\Delta
\log{L_{\rm Ly\alpha}}$, $\Phi^*$ is the normalisation, $L^*$ is the
characteristic break luminosity, and $\alpha$ is the slope of the
luminosity function. As our binned data-points are few, we choose not
to fit the Schechter function with three free parameters. Therefore we
assume $\alpha=-1.60$: the slope of the Schechter function for
$z\approx3$ Lyman Break Galaxies \citep{Steidel99}, which also fits
well to the luminosity distribution of $z\approx3$ Ly$\alpha$ emitters
\citep{Steidel00}. We determine the best fit by carrying out a
least-squares minimisation using $L^*$ and $\Phi^*$ as free
parameters. The best fit to our data is given by $L^* =
5.0\pm1.8\times10^{35}$~W and $\Phi^* = 0.0012\pm0.0005$~Mpc$^{-3}$
(shown in Fig.~\ref{discusplot} by the solid curve). To ensure
completeness we also fit a line to a dataset comprising our two
highest-luminosity points and the results from CH98 and Ku00 at $L> 5
\times 10^{35}$~W, which are at similar redshift. We then obtain $L^* =
5.4\pm1.7\times10^{35}$~W and $\Phi^* = 0.0022\pm0.0012$~Mpc$^{-3}$,
shown by the dashed curve in Fig.~\ref{discusplot}. The two fits
therefore agree with each other within the uncertainties.

It appears that the results of the higher-redshift surveys are largely
consistent with the lower-redshift Ly$\alpha$ luminosity function,
although the lowest-luminosity results are hampered by
incompleteness. AJ03 have inferred the same fact by comparing their
luminosity function with the results of CH98. Hu04 also report that
the properties of their $z=5.7$ sample shows the same properties as
their sample at $z\approx3$ \citep{Hu98}. The inferred conclusion is
that the luminosity function of Ly$\alpha$ emitters does not
significantly change from $z\approx3.4$ to $z=5.7$. \citet{Malhotra04}
have considered the evolution of the \Lya luminosity function at even
higher redshift, comparing samples at $z=5.7$ and $z=6.5$. They
conclude the luminosity functions do not statistically deviate
from each other, implying that the reionisation of the Universe must
have been largely complete at $z=6.5$.

We can calculate how many high-luminosity sources we would expect
 to find according to our luminosity function. In this way we can
 check our earlier statement that we find none of these objects
 because our probed volume is too small. If we integrate the
 luminosity function from the edge of our highest luminosity bin to
 infinity over the volume probed by our survey, we find that we would
 expect to find only one object with a higher Ly$\alpha$
 luminosity. The lack of high-luminosity sources in our survey is
 therefore within the Poisson error of our expectation and is not
 significant.

\subsection{The star-formation rate}

\begin{figure*}
\begin{center}
\includegraphics[width=16cm]{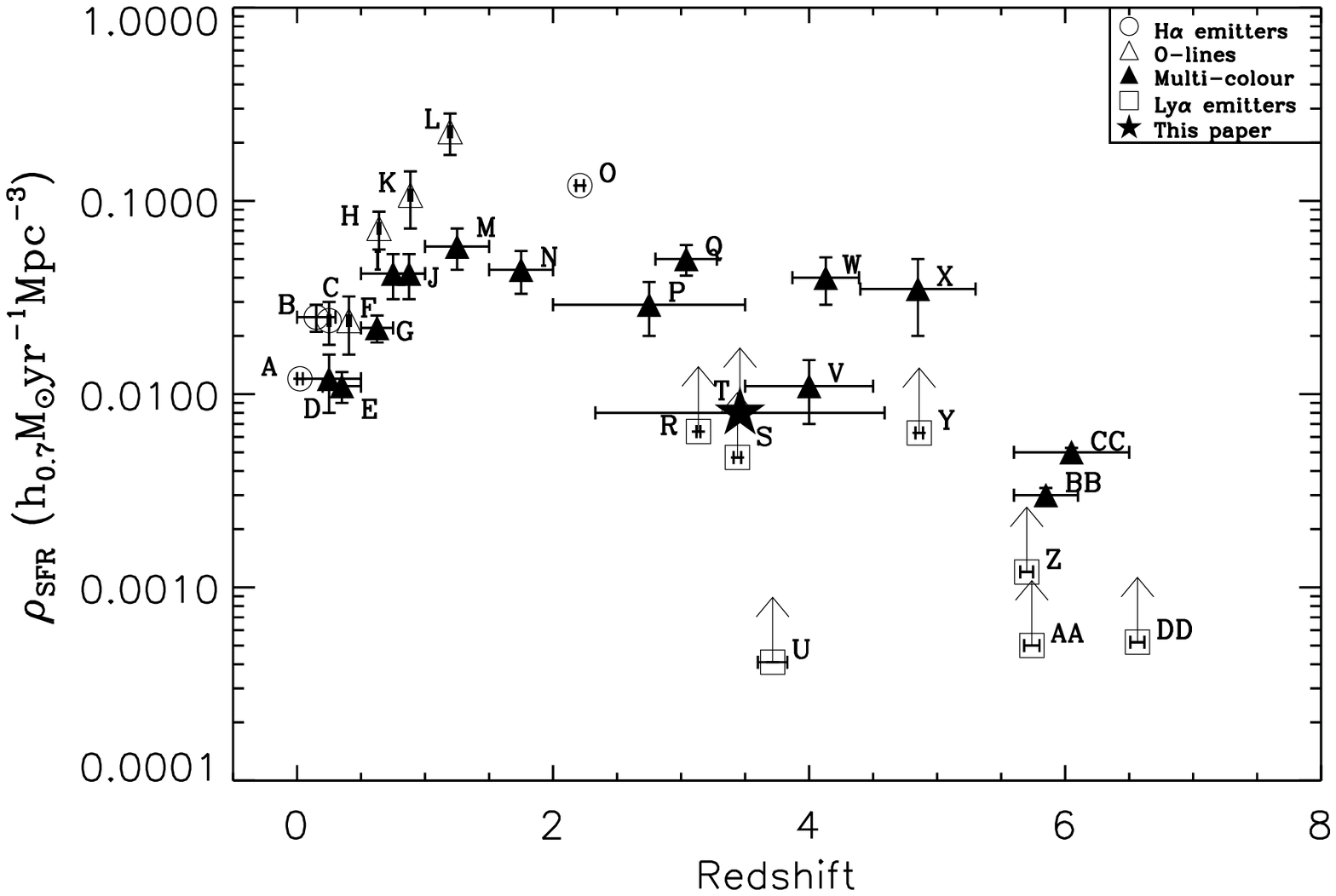}
\caption{\small Star-formation rate densities as derived by various
types of surveys. The result from this work derived from integrating
over the luminosity function fit to our data alone is denoted by the
star symbol.  The different types of surveys are marked with different
symbols: the open circles are H$\alpha$ searches, the open triangles
are surveys aimed at oxygen emission lines, the filled triangles are
multi-colour surveys, and the open squares are Ly$\alpha$ searches. All
the SFR densities and their relevant references can be found in
Table~\ref{labeltable}.  The x-error bars reflect the redshift range
of the surveys.}
\label{sfr}
\end{center}
\end{figure*}

As the Ly$\alpha$ line is excited when interstellar gas is
photoionised by the UV radiation of young and hot blue stars, the
Ly$\alpha$ luminosity of a galaxy is directly related to its star-formation
rate (SFR). We can derive this relation by using the case B recombination
theory \citep{Brocklehurst71} to deduce the H$\alpha$ luminosity of a
star-forming galaxy from its Ly$\alpha$ luminosity, and the equation found by
\citet{Kennicutt98} to link in turn its star-formation rate to its
H$\alpha$ luminosity. The formulae are respectively $L_{\rm Ly\alpha}
= 8.7L_{\rm H\alpha}$, and $SFR = 7.9 \times 10^{-35} L_{\rm H\alpha}$
M$_{\odot}$yr$^{-1}$, where $L_{\rm H\alpha}$ is the H$\alpha$
luminosity in Watts and continuous star formation is assumed with a
Salpeter Initial Mass Function from $0.1-100$~M$_{\odot}$. Thus we
get:

\begin{displaymath}
SFR=9.1 \times 10^{-36}\Big(\frac{L_{\rm Ly\alpha}}{\rm W}\Big)\rm{~M_{\odot}yr}^{-1}.
\end{displaymath}

Using this relation we estimate the star-formation rate for each of
our objects. It is important to note that these star-formation rates
are calculated under the assumption that the Ly$\alpha$ line is solely
photoionised by hot stars and neglecting any extinction. The SFRs
range from about 1 to 6 solar masses per year, with an average of $2.6
\pm 1.6$ M$_{\odot}$yr$^{-1}$.

Summing the individual SFRs of our objects and dividing by their
corresponding volume gives a comoving star-formation rate density of
$\rho_{SFR} = 0.0067 \pm
0.0005$~M$_{\odot}$yr$^{-1}$Mpc$^{-3}$. Integrating over the
luminosity function yields $\rho_{SFR} = 0.008
$~M$_{\odot}$yr$^{-1}$Mpc$^{-3}$ for $L^* = 5.0\times10^{35}$~W and
$\Phi^* = 0.0012$~Mpc$^{-3}$, which applies to our own dataset, and
$\rho_{SFR} \approx 0.017 $~M$_{\odot}$yr$^{-1}$Mpc$^{-3}$ for $L^* =
5.4\times10^{35}$~W and $\Phi^* = 0.0022$~Mpc$^{-3}$ as derived from
the combination of our dataset together with CH98 and Ku00. We compare
this SFR density to those from previous studies, compiled by
\citet{Kodaira03}, recalculated with our cosmological
parameters and more recent surveys (Fig.~\ref{sfr}, Table~\ref{labeltable}).

\begin{table*}
\centering
\caption{\small The star-formation rate densities found by previous
studies, shown in Fig.~\ref{sfr}.}
\begin{tabular}{c l l l l c l l l l}
\hline \hline
Label & $z$ & $\rho_{SFR}$ & Survey Type & Source & Label & $z$ & $\rho_{SFR}$ & Survey Type & Source  \\
& & $M_{\odot}yr^{-1}$ & & & & & $M_{\odot}yr^{-1}$ & & \\ 
\hline
A & 0.02 & 1.2$\times 10^{-2}$ & H$\alpha$    & \citet{Gallego96}  & P & 2.75 & 2.9$\times 10^{-2}$ & Multi-colour & \citet{Madau98}\\   
B & 0.15 & 2.5$\times 10^{-2}$ & H$\alpha$    & \citet{Tresse98}   & Q & 3.04 & 5.0$\times 10^{-2}$ & Multi-colour & \citet{Steidel99}\\ 
C & 0.25 & 2.4$\times 10^{-2}$ & H$\alpha$    & \citet{Hippelein03}& R & 3.14 & 6.4$\times 10^{-3}$ & Ly$\alpha$   & \citet{Kudritzki00}\\
D & 0.25 & 1.2$\times 10^{-2}$ & Multi-colour & \citet{Treyer98}   & S & 3.44 & 4.7$\times 10^{-3}$ & Ly$\alpha$   & \citet{Cowie98}\\   
E & 0.35 & 1.1$\times 10^{-2}$ & Multi-colour & \citet{Lilly96}    & T & 3.45 & 8.0$\times 10^{-3}$ & Ly$\alpha$   & This paper\\	  
F & 0.40 & 2.4$\times 10^{-2}$ & [OIII]	      & \citet{Hippelein03}& U & 3.72 & 4.1$\times 10^{-4}$ & Ly$\alpha$   & \citet{Fujita03}\\  
G & 0.63 & 2.2$\times 10^{-2}$ & Multi-colour & \citet{Lilly96}    & V & 4.00 & 1.1$\times 10^{-2}$ & Multi-colour & \citet{Madau98}\\   
H & 0.64 & 7.2$\times 10^{-2}$ & [OIII]	      & \citet{Hippelein03}& W & 4.13 & 4.0$\times 10^{-2}$ & Multi-colour & \citet{Steidel99}\\ 
I & 0.75 & 4.2$\times 10^{-2}$ & Multi-colour & \citet{Connolly97} & X & 4.85 & 3.5$\times 10^{-2}$ & Multi-colour & \citet{Iwata03}\\   
J & 0.88 & 4.2$\times 10^{-2}$ & Multi-colour & \citet{Lilly96}    & Y & 4.86 & 6.3$\times 10^{-3}$ & Ly$\alpha$   & \citet{Ouchi03}\\   
K & 0.88 & 1.1$\times 10^{-1}$ & [OII]	      & \citet{Hippelein03}& Z & 5.70 & 1.2$\times 10^{-3}$ & Ly$\alpha$   & \citet{Ajiki03}\\   
L & 1.20 & 2.3$\times 10^{-1}$ & [OII]	      & \citet{Hippelein03}& AA& 5.74 & 5.0$\times 10^{-4}$ & Ly$\alpha$   & \citet{Rhoads03}\\  
M & 1.25 & 5.8$\times 10^{-2}$ & Multi-colour & \citet{Connolly97} & BB& 5.80 & 0.3$\times 10^{-2}$ & Multi-colour & \citet{Stanway03}\\ 
N & 1.75 & 4.4$\times 10^{-2}$ & Multi-colour & \citet{Connolly97} & CC& 6.00 & 0.5$\times 10^{-2}$ & Multi-colour & \citet{Bunker04} \\ 
O & 2.21 & 1.2$\times 10^{-1}$ & H$\alpha$    & \citet{Moorwood00} & DD& 6.57 & 5.2$\times 10^{-4}$ & Ly$\alpha$   & \citet{Kodaira03}\\ 
\hline \hline
\end{tabular}
\label{labeltable}
\end{table*}

Although our result is lower than those of multi-colour surveys, it is
in good agreement with the Ly$\alpha$ surveys carried out around the
same redshift. The discrepancy between the Ly$\alpha$ searches
and the multi-colour surveys is probably because the Ly$\alpha$
searches only set a lower limit to the SFR density. This is due to the
fact that the SFR based on Ly$\alpha$ luminosity does not take account
of extinction by dust and interstellar and intergalactic gas. This is
a very important effect as Ly$\alpha$ is a resonant line and the
luminosity of the Ly$\alpha$ line can be severely affected by
absorption from H{\footnotesize{I}}. However, correcting for
obscuration requires extensive knowledge of the amount of dust and gas
in the light path and is therefore very uncertain and deep optical and
near-infrared observations are needed to make these corrections. To
compare our result with the multi-colour surveys, we can apply a
correction for the difference between the UV continuum emission and
the \Lya emission, which alleviates some of the problems associated
with H{\footnotesize{I}} absorption of the \Lya line.
We use $SFR(Ly\alpha) = 0.77 \times SFR(UV)$ \citep[see][]{Hu02}, where
$SFR(UV)$ is the star-formation rate estimate derived from the UV
continuum emission.
The corrected SFR density then becomes $\rho_{SFR} =
0.010$~M$_{\odot}$yr$^{-1}$~Mpc$^{-3}$ for our data and $\rho_{SFR} =
0.022$~M$_{\odot}$yr$^{-1}$~Mpc$^{-3}$ for the combined dataset. These
agree well with the SFR densities found by multi-colour surveys
at similar redshift, although we note the dust obscuration may still
be a severe problem. 

The lack of any evidence for evolution of the luminosity function of
\Lya emitters between $z\approx3.4$ and $z=5.7$ suggests that the
star-formation rate density does not change in this redshift interval,
although our survey is severely limited by low number
statistics. Inspection of Fig.~\ref{sfr} reveals that various surveys
around $z=6$ show evidence for a decrease in the cosmic star-formation
rate density toward high redshift. For Ly$\alpha$-emitter surveys this could
be caused by the increasing optical depth of H{\footnotesize{I}} to
the \Lya emission line. However, the multi-colour surveys
\citep{Stanway03, Bunker04} which calculate the cosmic star-formation
rate density by integrating over the luminosity function of Lyman
Break Galaxies, also point to a decrease in the star-formation
density. A future wider-field IFU survey would result in a large
sample of \Lya emitters discovered in a perfectly consistent
manner. This would facilitate investigation of the
star-formation rate density within a large redshift range by comparing
exactly the same sort of objects with the same selection
criteria. This could provide significant clues to the evolution of the
star-formation rate density throughout the Universe.

\subsection{Clustering around the radio galaxy}
It is interesting to determine whether the fact that our observations
were centred on radio galaxy MRC 0943-242 at $z=2.92$ has any
influence on the number of \Lya emitters we find. \citet{Venemans02}
suggest that luminous radio galaxies are tracers of regions of galaxy
overdensities in the early Universe. They find \Lya emitters around
the radio galaxy TN J1338-1942 at $z=4.1$ with a relative overdensity
of $\sim 15$ compared to blank-field surveys. \citet{Kudritzki00}
carried out a \Lya search with a similar flux limit to our average
limit, targeted at a blank field at $z=3.1$ which is close to the
redshift of our radio galaxy. They derive a number density of $14400
\Delta$z$^{-1}$deg$^{-2}$.  Based on this we would expect to find
$\sim 1$ \Lya emitter at redshift $z
\sim 2.92$ over our entire field, within a redshift interval of
$\Delta z \sim 0.004$. Indeed we find one emitter with $z_{\rm rg} -
z_{\rm em} = 0.002$ (object 5644).

It is evident that the angular size of our field is
much too small to determine whether there is an overdensity around our
target. Although it shows that we can compare our results with blank field
studies without a significant bias due to the presence of the radio
galaxy.

We also note that three of our \Lya emitters lie withing $\Delta z =
0.04$ of each other at $z \sim 2.5$. It is difficulty to say that this
is definitely an overdensity, but it is suggestive and further
observations are needed. However, this shows the power that
IFU-surveys may have in identifying regions of galaxy overdensities at
high redshift.

\section{SUMMARY}
We summarise our results and conclusions as follows:
\begin{itemize}
\item We have discovered 14 Ly$\alpha$ emitters with a flux varying from
$1.4$ to $7.3 \times 10^{-20}$ W m$^{-2}$ in a comoving volume of
$7.8\times10^3$Mpc$^3$. The resulting number density is
$0.0018^{+0.0006}_{-0.0005}$ Mpc$^{-3}$.

\item Fitting a Schechter luminosity function to the data shows that
the luminosity distribution of the sources is comparable to that of
Ly$\alpha$ emitters at $z=5.7$. This implies that we do not see any
evolution of Ly$\alpha$ emitters between $z \approx 3.4$ and $z =
5.7$.

\item The star-formation rate of the individual objects lies within a
range from about 1 to 6 solar masses per year. The derived cosmic star
formation rate density is
$\rho_{SFR}=6.7\pm0.5\times10^{-3}$~M$_{\odot}$yr$^{-1}$Mpc$^{-3}$. This
is consistent with the results of Ly$\alpha$ emitter searches at
similar redshift. It is however significantly lower than the results
of multi-colour surveys. This is due to the fact that the Ly$\alpha$
searches only yield a lower limit to the cosmic star-formation rate
density because of absorption of the Ly$\alpha$ emission and the lack
of a reliable luminosity function.  We can compare our result to the
multi-colour surveys if we determine the cosmic star-formation rate by
integrating our fit of the Schechter luminosity function to the
combination of our dataset with surveys at similar redshift and
correcting for the difference of extinction between Ly$\alpha$
luminosity and UV luminosity. We then obtain $\rho_{SFR} \sim
2.2\times10^{-2} $~M$_{\odot}$yr$^{-1}$Mpc$^{-3}$, which is similar to
multi-colour results.

\item We find one companion to the central radio galaxy at $z = 2.92$. We do not find
further evidence for clustering around the radio galaxy, as our field
of view is too small to allow for such investigations. However, we do
find three \Lya emitters within $\Delta z = 0.04$ at $z \sim 2.5$
which suggests that there may be an overdensity of galaxies at this
redshift, although further observations are needed to confirm this.

\item An important conclusion to this project is that the IFU-technique is
well suited to detect Ly$\alpha$ emitters and trace the star-formation
history of the Universe. It is however vital to carry out larger
surveys to probe a larger volume and therefore detect a higher number
of sources. It will then become possible to reliably sample the
luminosity density of Ly$\alpha$ emitters throughout a large redshift
range with a single survey. This will facilitate studying the
evolution of the properties of Ly$\alpha$ emitters throughout time and
therefore map the evolution of the cosmic star-formation rate
density reliably, through comparison of galaxies selected by exactly
the same method.
\end{itemize}

\section*{ACKNOWLEDGMENTS} 
We thank Wil van Breugel for the B-, V- and I-band images and the LRIS
spectrum. We would also like to thank Steve Rawlings, Huub
R\"ottgering and Richard Wilman for useful discussions. The IFU data
published in this paper have been reduced using VIPGI, developed by
INAF Milano, in the framework of the VIRMOS Consortium activities. We
thank Bianca Garilli and Marco Scodeggio for their help regarding
VIPGI. CVB would like to acknowledge the financial support from the
ERASMUS program, the International Study Fund of Leiden University
(LISF) and a PPARC studentship, MJJ acknowledges the support of a
PPARC PDRA.



\end{document}